\documentclass[a4paper,twocolumn,11pt,unpublished]{quantumarticle}
\pdfoutput=1
\usepackage[utf8]{inputenc}
\usepackage[english]{babel}
\usepackage[T1]{fontenc}
\usepackage{amsmath,amssymb}
\usepackage{hyperref}
\usepackage[numbers,sort&compress]{natbib}
\usepackage{braket}
\usepackage{comment}
\usepackage{algorithm}
\usepackage{algpseudocode}
\usepackage{float}

\begin{document}
\title{Scheduling of syndrome measurements with a few ancillary qubits}

\author{Shintaro Sato}
\email{shintaro.sato@ntt.com}
\affiliation{NTT Computer and Data Science Laboratories, NTT, Inc., Musashino 180-8585, Tokyo, Japan}

\author{Yasunari Suzuki}
\affiliation{NTT Computer and Data Science Laboratories, NTT, Inc., Musashino 180-8585, Tokyo, Japan}
\orcid{0000-0002-8005-357X}
\maketitle

\begin{abstract}
Quantum error-correcting codes are a vital technology for demonstrating reliable quantum computation.
They require data qubits for encoding quantum information and ancillary qubits for taking error syndromes necessary for error correction.
The need for a large number of ancillary qubits is an overhead specific to quantum computing, and it prevents the scaling of quantum computers to a useful size.
In this work, we propose a framework for generating efficient syndrome measurement circuits with a few ancillary qubits in CSS codes and provide a method to minimize the total number of physical qubits in general settings.
We demonstrated our proposal by applying it to surface codes, and we generated syndrome measurement circuits under several constraints of total qubit count.
As a result, we find that balanced data and ancillary qubit counts achieve the lower logical error rates under a fixed total number of physical qubits.
This result indicates that using fewer ancillary qubits than the number of stabilizers can be effective for reducing logical error rates in a practical noise model.
\end{abstract}

\section{Introduction}\label{sec:Introduction}
Quantum error correction~(QEC) is a vital technology for performing large-scale and fault-tolerant quantum computing in practical applications.
In the QEC process, logical qubits are represented by several physical qubits called data qubits, and errors are detected via syndrome measurements. These measurements can be performed with indirect Pauli measurements using ancillary physical qubits.
It is known that if the physical error rate is smaller than a threshold value, the error rate of logical qubits can be reduced to an arbitrarily small value by increasing the number of data qubits per logical qubit~\cite{shor1996fault,knill1996concatenated,knill1998resilient, aharonov1997fault}. 
In recent years, there has also been progress in demonstrating the utilities of QEC codes using actual devices~\cite{acharya2022suppressing,acharya2024quantum,bluvstein2024logical,reichardt2024demonstration}.

A major difficulty in implementing QEC codes is that a large number of physical qubits are required to construct logical qubits.
One reason why QEC codes need many qubits is that we typically use the same number of ancillary qubits as the syndrome measurements to perform them in high parallelism~\cite{fowler2012surface,fowler2018low}.
This means only a part of the physical qubits can be used as data qubits, which reduces the number of data qubits available for a logical qubit and limits the performance of QEC.

A promising approach to mitigating this problem is to perform syndrome measurements with fewer ancillary qubits by reusing them during the syndrome measurement process. 
Refs.\,~\cite{PhysRevLett.110.070503, PhysRevA.107.032403} proposed methods to perform syndrome measurements with a single ancillary qubit for specific graph structures and error-correcting codes. Ref.\,\cite{ye2025quantumerrorcorrectionlong} considers the error correction with a few ancillary qubits on the device with all-to-all connectivity.
On the other hand, there are two problems with the existing approaches. 
First, existing approaches cannot be applied to general situations, such as different QEC codes, ancillary qubit counts, and qubit connectivities. 
Second, the depth of the syndrome measurement circuits is significantly increased as a trade-off for reducing the number of ancillary qubits.
This is a crucial drawback since deeper quantum circuits incur more accumulated idling errors and result in higher logical error rates. 
If the circuit depth is too long, more data qubits per logical qubit are needed to maintain logical error rates. Thus, ancillary-qubit reuse might increase the total number of physical qubits rather than decrease it.
To address these issues, we require a general framework for generating syndrome measurement circuits with fewer ancillary qubits. The number of ancillary qubits should be tuned to minimize the total physical qubit count while achieving the target logical error rates. 
However, this point has not been explored yet.

In this paper, we propose a framework for generating efficient syndrome measurement circuits with a few ancillary qubits in general cases and evaluate the necessary number of total physical qubits in several settings.
In our framework, we formalize a state of the syndrome measurement process and model the possible transitions via two-qubit gates. 
This framework can generate an efficient syndrome measurement circuit by searching for a short transition sequence to a state in which all the syndrome measurements are finished. 
We also provide a greedy-based algorithm to find such a sequence. 
By appropriately classifying the syndrome measurement status, our algorithm efficiently reuses ancillary qubits while avoiding getting stuck.
Since our formalism and algorithm are defined in a general manner, our method can be applied to any situation as long as the target QEC code is a CSS code.

We demonstrated our algorithm by generating syndrome measurement circuits for surface codes and varied the ratio of ancillary qubits. 
When the code distance is fixed, we observe that the logical error rates decrease as the number of ancillary qubits increases, particularly when idling errors are dominant. 
If we fixed the number of available physical qubits, we observed that logical error rates are minimized when the number of data and ancillary qubits is appropriately balanced. 
Our results indicate that choosing the number of ancillary qubits to be less than the number of syndrome measurements is effective in minimizing logical error rates under a fixed device size. In other words, our methods can reduce the number of physical qubits required to achieve the desired logical error rates.

Our contributions in this paper can be summarized as follows:
\begin{itemize}
    \item We formulated a problem of finding an efficient syndrome measurement circuit with a few ancillary qubits in the general settings and provided an algorithm to generate it.
    \item We benchmarked the performance of our methods with surface codes and revealed the relation between logical and physical error rates under various types of noise models.
    \item We evaluated the optimal number of ancillary qubits using our framework under a fixed total number of physical qubits. The results indicate that balanced data and ancillary qubit counts achieve lower logical error rates.
\end{itemize}

The outline of the paper is as follows. 
In Section~\ref{sec:Preliminaries}, we explain the basics of syndrome measurements of QEC for stabilizer codes and give notations for them.
Section~\ref{sec:Scheduling} introduces our framework for designing an efficient syndrome measurement circuit and defines scheduling problems with it.
Section~\ref{sec:Algorithm} provides an algorithm to solve the scheduling problems.
In section~\ref{sec:Numerical}, we numerically evaluate the performance of our method with rotated surface code on a two-dimensional grid.
In section~\ref{sec:Conclusion}, we discuss the availability of our proposal and future directions.
The appendices provide details of our algorithms.

\section{Basics of syndrome measurements}\label{sec:Preliminaries}
This section provides a brief introduction to the basics of stabilizer codes and introduces the key notions used in this paper.
The code space of stabilizer codes can be defined as the simultaneous $+1$ eigenspace of a commutative and independent subset of Pauli operators, which is called a stabilizer generator set. CSS stabilizer codes are a class of stabilizer codes, of which the stabilizer generator set consists of Pauli-$X$ operators and Pauli-$Z$ operators. Many popular QEC codes are CSS codes, and this paper focuses on the implementation of CSS stabilizer codes.
When we protect logical qubits with stabilizer codes, we need to perform Pauli measurements with each Pauli operator in the stabilizer generator set, obtain syndrome values, and estimate physical errors based on these values.

\begin{figure*}[t]
  \centering
  \begin{tabular}{c|c}
    \includegraphics[width=0.5\textwidth]{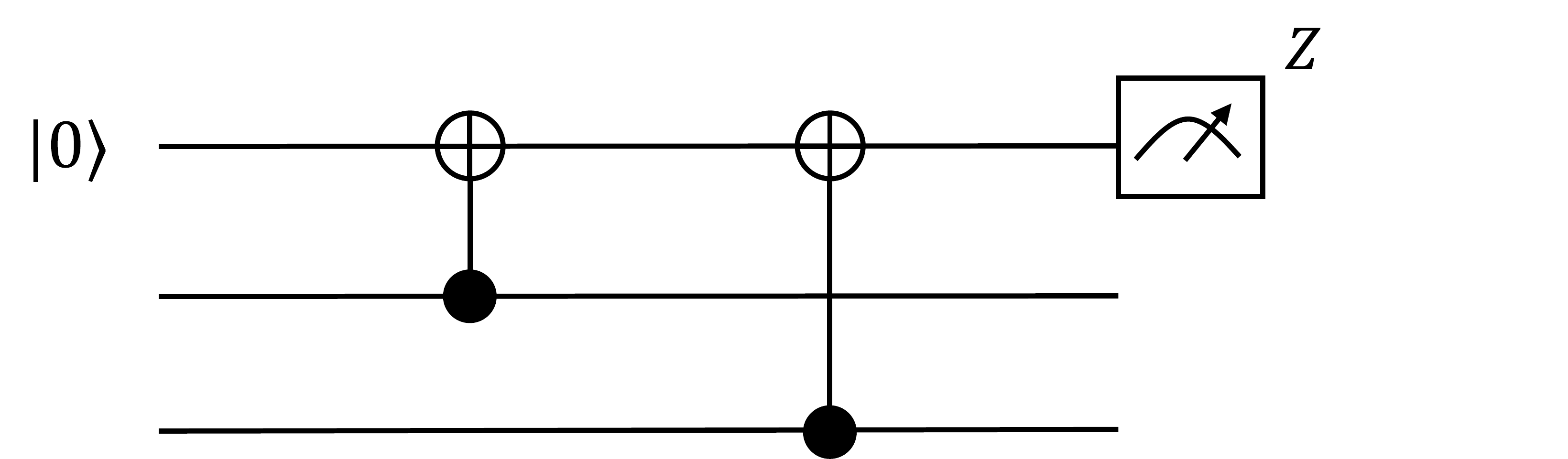} &  \includegraphics[width=0.5\textwidth]{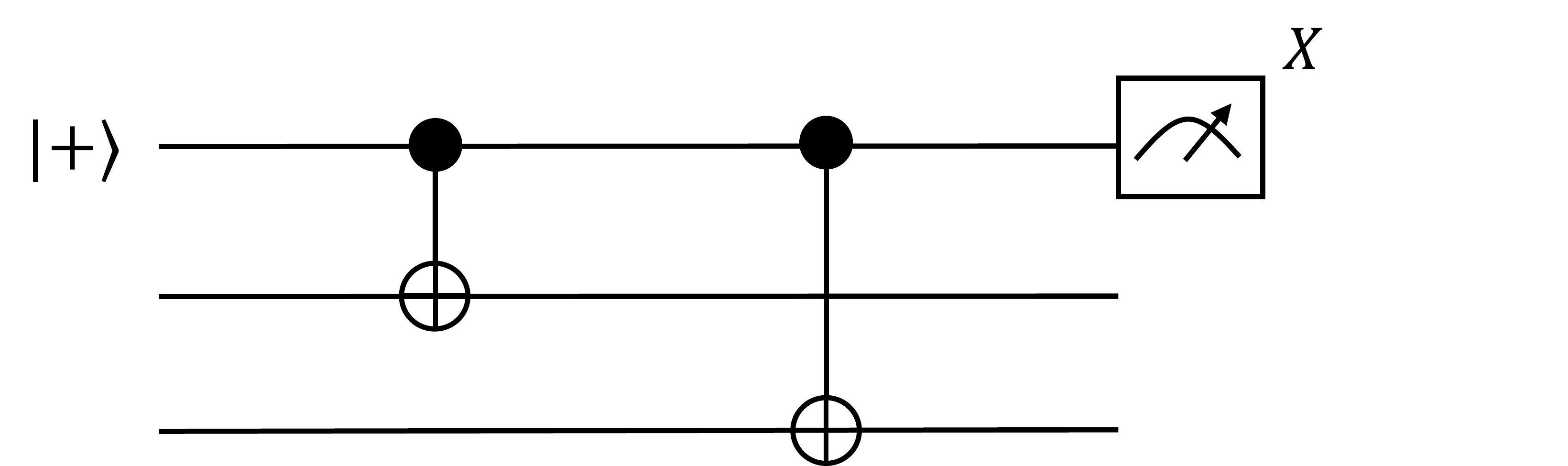}
  \end{tabular}
      \caption{(Left) : Syndrome measurement circuit for $Z^{\otimes2}$. (Right) : Syndrome measurement circuit for $X^{\otimes2}$}
      \label{fig:meas_sample}
\end{figure*}

The Pauli measurements for obtaining syndrome values are often performed via indirect Pauli measurements with an ancillary qubit, as shown in Fig.\,~\ref{fig:meas_sample}.
For example, we can perform a multi-qubit Pauli-$Z$ measurement acting on the subset of data qubits by initializing an ancillary qubit to $\ket{0}$, performing CNOT gates from each target data qubit, and measuring the ancillary qubit in the Pauli-$Z$ basis. To ensure the parallelism in syndrome measurements, most existing implementations use an independent ancillary qubit for each Pauli measurement. In this case, the same number of ancillary qubits as the size of the stabilizer generator set is needed in addition to the data qubits.

In this paper, we denote the number of data and ancillary qubits as $n$ and $m$ and label them with $(d_1,\cdots, d_n)$ and $(a_1, \cdots, a_m)$, respectively. The operators $X_i$ and $Z_i$ denote the Pauli-$X$ and $Z$ operators acting on the data qubit $d_i$, respectively.
We assume qubit connectivity is represented by a connected graph $G=(V, E)$, where $V=(v_1, \cdots, v_{n+m})$ is the location of qubit devices, and two-qubit gates can be performed if the target qubits are located on the vertices connected by edge $e \in E$.
Since we focus on CSS stabilizer codes, we denote a set of Pauli-$X$ ($Z$) operators in the stabilizer generator set as $G_X$ ($G_Z$).

\section{Scheduling problem of syndrome measurement circuits}\label{sec:Scheduling}
To effectively utilize a given number of ancillary qubits in syndrome measurements, it is necessary to develop a framework that tracks the syndrom-extraction process and an algorithm for generating the optimized syndrome-extraction circuit.

In this section, we introduce our formulation of scheduling problems of syndrome extraction circuits. We first illustrate our description with an example of the repetition code in a one-dimensional graph, and then we show a generalized description. An algorithm for solving this scheduling problem is presented in the next section.

\begin{figure*}[!htbp]
  \centering
  \includegraphics[width=1.0\textwidth]{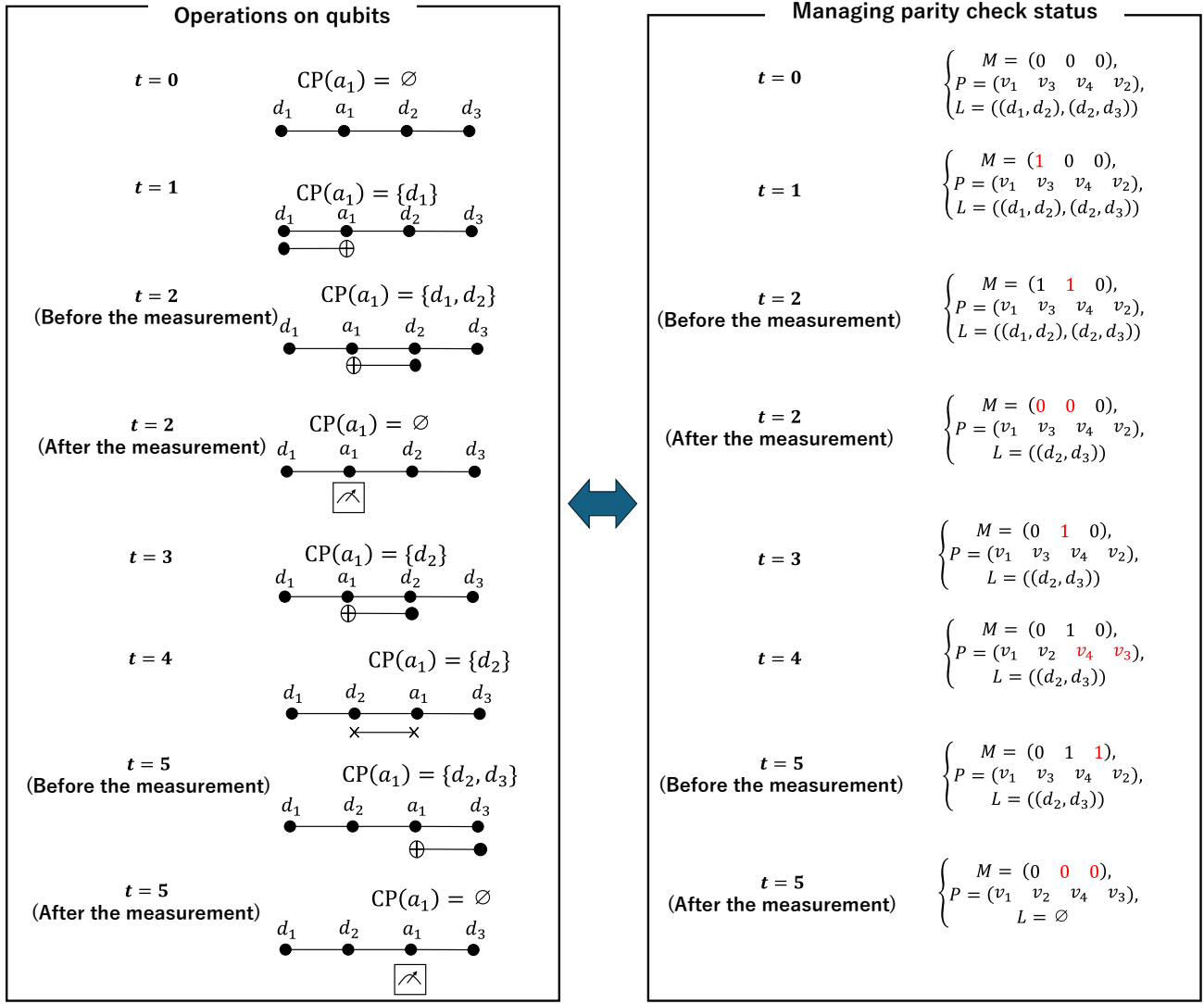}
  \caption{The left figure shows an example procedure of syndrome extraction for 1D repetition codes on 1D-array qubits. The right figure shows the corresponding state descriptions of syndrome extraction in our framework. The changes from the previous step are highlighted in red. The bottom quantum circuit shows the corresponding syndrome extraction circuit.}
  \label{fig:managing}
\end{figure*}

We consider the case where we perform the syndrome measurement of a distance-$3$ repetition code, of which the stabilizer generator set is $G_Z = \{Z_1 Z_2, Z_2 Z_3\}$. 
We assume one-dimensional qubit connectivity and initial qubit locations as shown in the left of Fig.\,~\ref{fig:managing}.
For simplicity, in our framework, we assume that the time for single-qubit operations is negligible and only consider the depth of two-qubit gates, i.e., CNOT and SWAP gates.
We assume ancillary qubits are initialized to $\ket{0}$ and obtain the syndrome values with CNOT gates and Pauli-$Z$ measurements. If ancillary qubits and target qubits are not neighboring, we can use SWAP gates to make them neighbors.
The left of Fig.\,~\ref{fig:managing} shows an example procedure; two CNOT gates and a single-qubit measurement are performed for Pauli-$Z_1 Z_2$ measurement (from $t=0$ to $t=2$), and Pauli-$Z_2 Z_3$ follows that using SWAP gates (from $t=3$ to $t=5$).
In each time step, ${\rm CP}(a_1)$ denotes a set of data-qubit labels to which CNOT gates are applied from the ancillary qubit $a_1$. 

We track the status of the syndrome measurement process with the following three variables: 
\begin{enumerate}
\item Parity check processing matrix $M \in \{0,1\}^{m \times n}$
\item Qubit-to-localtion map $P \in {\rm Perm}(V)$
\item Unmeasured operator label list $L$
\end{enumerate}
Note that ${\rm Perm}(V)$ is the set of all the permutations of node list $V$.
Parity check processing matrix $M$ is a $m \times n$ binary matrix, and $M_{i,j} = 1$ indicates that $d_j \in {\rm CP}(a_i)$. In other words, $M_{i,j}=1$ if a CNOT gate on the ancillary qubit $a_i$ and data qubit $d_j$ is performed for checking the parity of $d_j$.
A qubit-to-location map $P$ represents locations of qubits. The $i$-th data qubit $d_i$ is at the location $P_i$, and the $i$-th ancillary qubit $a_i$ is at the location $P_{n+i}$. An unmeasured operator label list $L$ is a set of data-qubit label lists of unmeasured stabilizer generators. For example, a variable $L$ corresponds to $G_Z = \{Z_1 Z_2, Z_2 Z_3\}$ is $L=((d_1, d_2), (d_2, d_3))$. The state of syndrome extraction is represented by the tuple $(M,P,L)$. The correspondence of the 1D syndrome measurement process is represented on the right of Fig.\,\ref{fig:managing}.

Using this description, we can formulate the initial state, update rules, and final state as follows.
Suppose we construct a syndrome measurement circuit for Pauli-$Z$ stabilizers. At the initial state, $M$ is a zero matrix and $L$ is equal to the stabilizer generator set $G_Z$. We assume $P$ is provided as a part of the problem instances.
If we perform CNOT gates from ancillary qubit $a_i$ to data qubit $d_j$, the element of $M_{i,j}$ will be flipped. When we perform SWAP gates on a pair of neighboring qubits, two elements in $P$ will be swapped. If the $i$-th row of $M$ corresponds to the indices of an unmeasured label list $l \in L$, a single-qubit Pauli-$Z$ measurement on $a_i$ can be performed and removes $l$ from $L$ and resets the $i$-th row of $M$ to a zero vector. When $L$ becomes empty, the syndrome measurement is completed. 

Note that if the matrix $M$ is not a zero matrix at the final state, ancillary qubits are entangled with data qubits. This can be safely disentangled by measuring the ancillary qubit in the $X$ basis, but such a situation can be eliminated by optimization, as explained later.

By using the above formalism, the problem of constructing efficient syndrome-extraction circuits can be rephrased as the following scheduling problem:
\textit{For given qubit connectivity $G$, initial qubit locations $P$, and stabilizer generator set $L$, is there an efficient algorithm that makes $L$ empty with a small number of updates?}
We will provide a scheduling algorithm for this problem in the next section.

Note that the above problem is a problem to build quantum circuits for the Pauli-$Z$ stabilizer generator set, and we also need to perform the same operation for the Pauli-$X$ stabilizer generator set. This can be achieved as follows.
Suppose we find a sequence of one- and two-qubit gates $S_Z$ for $G_Z$. If the target stabilizer code is self-dual, i.e., the indices of Pauli operators in $G_X$ are equal to $G_Z$, we can perform the Pauli-$X$ stabilizer measurement with the reversed sequence $S_Z^{-1}$ by replacing the basis of $X$ and $Z$.
Otherwise, we set the positions after the Pauli-$Z$ syndrome measurements as the initial locations $P$, solve the scheduling problem for $G_X$, and obtain a sequence $S_X$.
Then, we can construct one round sequence of syndrome measurements as $S_Z \rightarrow S_Z^{-1}\rightarrow S_X \rightarrow S_X^{-1}$, where $S_X^{-1}$ and $S_Z^{-1}$ are the reversed sequence of syndrome measurements.

\section{Scheduling algorithm}\label{sec:Algorithm}
This section presents an algorithm for searching efficient syndrome extraction circuits by solving the problem formulated in the previous section. We first provide an overview of our algorithm by assuming several functions in Sec.\,\ref{alg:overview}. Then, we describe the details of each function.

\subsection{Scheduling algorithm (Algorithm\,\ref{alg:main_alg})} \label{alg:overview}
The idea behind our algorithm is as follows: in each time step, the state of each ancillary qubit is classified into four cases based on two conditions. Then, we determine the action of each ancillary qubit according to the assigned case. We repeat this procedure until all the stabilizer measurements are completed.

The overview of our algorithm is shown in Algorithm\,\ref{alg:main_alg}. 
\begin{algorithm}[H]
\caption{Scheduling algorithm outline}
\label{alg:main_alg}
\begin{algorithmic}[]
\Require{$(P, L, G)$}
\State{$M \leftarrow 0^{m \times n}$}
\State{$C \leftarrow ()$}
\While{$L \neq \emptyset$}
    \State{$(C_t,M,P,L) \leftarrow \texttt{Alg}\,\ref{alg:classify} (M,P,L,G)$}
    \If{$C_t = ()$}
        \State{$(C_t,P) \leftarrow \texttt{Alg}\,\ref{alg:tie_brekaing}(M,P,L,G)$}
    \EndIf
    \State{$C \leftarrow C + C_t$}
\EndWhile
    \State{$C \leftarrow \texttt{Alg}\,\ref{alg:remove_gates}(C,M)$}
\State \textbf{return} $C$
\end{algorithmic}
\end{algorithm}
Given input data, this algorithms repeatedly call Algorithm\,\ref{alg:classify}, which determines the cases of ancillary qubits, writes actions of physical qubits at this timestep $t$ to $C_t$. We also update the status of the syndrom extraction $(M,P,L)$ according to the actions.
To avoid a stucking or infinite-loop situation, we invoke the tie-breaking Algorithm\,\ref{alg:tie_brekaing} if no action happens.
Then, the chosen actions are appended to the syndrom-extraction circuits $C$.
After the syndrome measurements are completed, we call Algorithm\,\ref{alg:remove_gates} to reduce the depth of the circuit by removing unnecessary gates.
The rest of this section will describe the Algorithms\,\ref{alg:classify},\,\ref{alg:tie_brekaing} and \,\ref{alg:remove_gates}.

Note that it is not trivial whether our scheduling always halts within a finite timestep for any stabilizer codes. In fact, we observed that several simple heuristic approaches easily tend to infinite loops. Our construction is carefully designed to guarantee that the algorithm halts within a finite time rather than aggressively optimizing schedules. See Appendix~\ref{app:halting} for the discussion of this property, and Sec.\,\ref{sec:Conclusion} for future directions.

\subsection{Decide Action (Algorithm\,\ref{alg:classify})}
This section explains Algorithm\,\ref{alg:classify} to classify ancillary qubit states and select gate operations.
\begin{algorithm}[H]
\caption{Algorithm to decide actions}
\label{alg:classify}
\begin{algorithmic}[]
    \Require{$(M, P, L, G)$}
    \State{$C_t \leftarrow ()$}\Comment{action list at this timestep}
    \State{$\texttt{flag} \leftarrow \texttt{true}$}\Comment{Flag for case2}
    \ForAll{$q \in (d_1 ... d_n, a_1 ... a_m)$}
        \State{$U[q] \leftarrow \texttt{true}$}
        \Comment{\texttt{true} if qubit unused}
    \EndFor
    \For{$i \in [1..m]$}
        \If{$\neg U[a_i]$}
            \State{\textbf{continue}}
        \EndIf
        \State{$\texttt{cond1} \leftarrow (\exists j \,\textrm{s.t.}\, M_{i,j} = 1$)}
        \State{$\texttt{cand} \leftarrow \texttt{get\_candidate}(M,P,L,G,U)$}
        \State{$\texttt{cond2} \leftarrow (\texttt{cand} \neq \texttt{None})$}
        \If{$\texttt{cond1} \And \texttt{cond2}$}
            \Comment{Case1}
            \State{$\texttt{cnot}(a_i,\texttt{cand},C_t,M,L,U)$}
        \EndIf
        \If{$\texttt{cond1} \And \neg \texttt{cond2}$}
            \Comment{Case2}
            \State{$\texttt{neighbor},\texttt{data} \leftarrow \texttt{get\_target}(M,P,L,G,U)$}
            \If{$\texttt{neighbor} \neq \texttt{None}$}
                \State{$\texttt{swap}(a_i,\texttt{neighbor},C_t,P,U)$}
                \If{$\texttt{flag} = \texttt{true}$}
                    \State{$U[\texttt{data}] \leftarrow \texttt{false}$}
                    \State{$\texttt{flag} \leftarrow \texttt{false}$}
                \EndIf
            \EndIf
        \EndIf
        \If{$\neg \texttt{cond1} \And \texttt{cond2}$}
            \Comment{Case3}
            \State{$\texttt{cnot}(a_i,\texttt{cand},C_t,M,L,U)$}
        \EndIf
        \If{$\neg \texttt{cond1} \And \neg \texttt{cond2}$}
            \Comment{Case4}
            \State{\textbf{pass}}
        \EndIf
    \EndFor
    \State \textbf{return} $C_t, M, P, L$
\end{algorithmic}
\end{algorithm}

This algorithm works as follows. First, all the physical qubits are marked as \texttt{unused} (described with a variable $U$ in pseudo-code). The \texttt{unused} flag indicates that no action is assigned to the qubit at the current time step. We classify the state of each unused ancillary qubit into four cases. Then, we determine which gates are assigned to them based on the classified case. This classification is performed based on whether the following two conditions are satisfied or not.
\begin{itemize}
\item \textbf{Condition 1:} The ancillary qubit $a_i$ has already interacted with any data qubit after the last initialization, i.e., at least one 1-element exists in the $i$-th row of $M$.
\item \textbf{Condition 2:} The ancillary qubit $a_i$ has neighboring data qubits with which interaction advances a syndrome measurement. Precisely saying, we call a data qubit $d_j$ is candidate if $d_j$ is marked as \texttt{unused}, the locations of $a_i$ and $d_j$ is neighboring, $d_j \notin {\rm CP}(a_i)$, and there is an element $l\in L$ such that ${\rm CP}(a_i) \cup \{d_j\} \subseteq l$. If there are multiple candidates, we pick the first one in a certain fixed order (e.g., pick the candidate with the smallest index). A function to get the candidate is described with \texttt{get\_candidate} in the pseudo-code. Condition 2 is satisfied if there is at least one candidate data qubit.
\end{itemize}

The four cases and their actions are as follows. Note that when actions are performed on physical qubits (i.e., when functions such as \texttt{cnot} and \texttt{swap} in the pseudo-code are called), the syndrome extraction state $(M,P,L)$ and unused qubit list $U$ are also updated.

\paragraph{Case 1:} An ancillary qubit $a_i$ is classified as case 1 if both conditions are satisfied. If the ancillary qubit is classified as case 1, we perform a CNOT gate on $a_i$ and any candidate data qubit. If there is $l \in L$ such that $l = {\rm CP}(a_i)$ after this operation, we perform the Pauli-$Z$ measurement on $a_i$.

\paragraph{Case 2:} 
An ancillary qubit $a_i$ is classified to case 2 if it satisfies Condition 1, but does not satisfy Condition 2. 
If the ancillary qubit is classified as case 2, a SWAP gate is performed to make $a_i$ close to a certain data qubit to complete the unfinished Pauli measurement. The target of the SWAP gate is determined as follows. We extract all the unfinished operators that might be finished by $a_i$ as $\{l \in L | {\rm CP}(a_i) \subset l\}$, and take the largest set among them. If there are multiple largest sets, we pick the first set in a certain ordering. Then, we choose the first data-qubit label in $l$ that is not measured by $a_i$, and perform SWAP gates on the qubit neighboring to $a_i$ to make $a_i$ close to the picked data qubit. We calculate the target data qubit in $l$ and the neighboring qubit for a SWAP gate with the function \texttt{get\_target} in the pseudo-code. If there is no such neighboring qubit, this function returns \texttt{None} and no action is assigned to the ancillary qubit.

Note that when we assign case 2 to the ancillary qubit for the first time in each time step, we mark the target data qubit \texttt{used} (i.e., $U[\texttt{data}]\leftarrow \texttt{false}$) in addition to the ancillary and neighboring qubits, which is controlled by a variable \texttt{flag} in the pseudo-code. This is for guaranteeing the halting property of our algorithm. See Appendix.\,\ref{app:halting} for details.

\paragraph{Case 3:}
An ancillary qubit is classified as case 3 if Condition 1 is not satisfied but Condition 2 is satisfied. If the ancillary qubit is classified to case 1, we perform a CNOT gate on $a_i$ and any candidate data qubit. Note that we assume each Pauli operator in the stabilizer generator set acts on more than one qubit, and no measurement occurs after the action of case 3.

\paragraph{Case 4:}
An ancillary qubit is classified as case 4 if neither condition is satisfied.
The case 4 ancillary qubits do nothing at this time step.

\begin{figure*}[!htbp]
  \centering
  \includegraphics[width=1.0\textwidth]{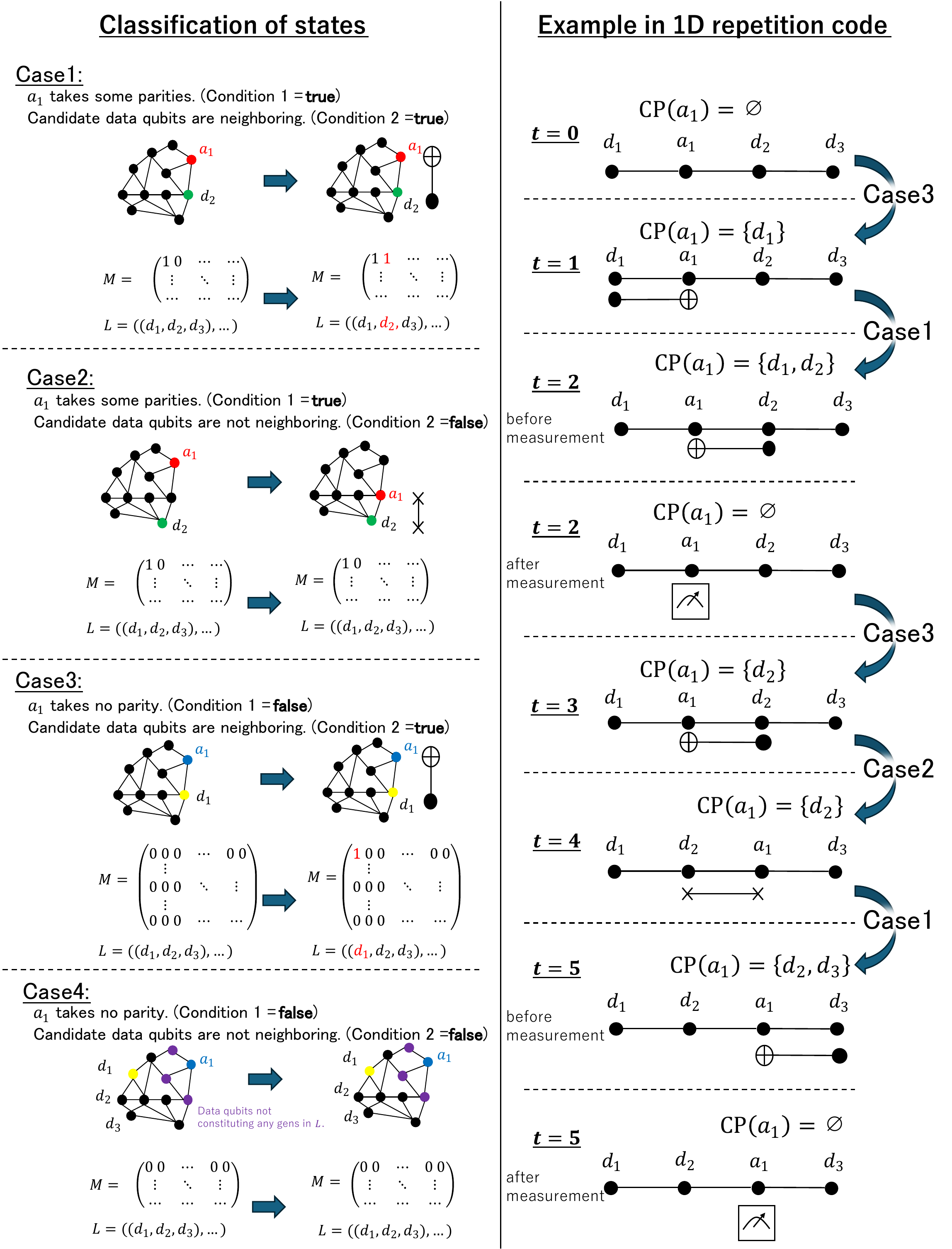}
  \caption{(Left) Four cases assigned in Algorithm.\,\ref{alg:classify}. (Right) The assignment of four cases to an ancillary qubit in the example of 1D repetition code shown in Fig.\,\ref{fig:managing}.}
  \label{fig:example_cases}
\end{figure*}

The four classes and their actions are depicted on the left of Fig.\,\ref{fig:example_cases}.

\subsection{Tie Break (Algorithm\,\ref{alg:tie_brekaing})}
In some situations, Algorithm\,\ref{alg:classify} assigns no action to every qubit and scheduling gets stuck. To avoid such situations, we invoke the tie-breaking algorithm to force qubit actions. 
Since the details of this algorithm are complicated, the detailed algorithm is explained in Appendix~\ref{app:alg_tie_break}. Here, we briefly describe the algorithm.
The Algorithm\,\ref{alg:classify} assigns no action to all the ancillary qubits if all the ancillary qubits are classified to case 4\footnote{Note that ancillary qubits with case 2 might be assigned no action when other ancillary qubits have already used neighboring qubits that the ancillary qubit tries to interact with. On the other hand, this indicates there is at least one ancillary qubit with an action. Thus, the tie-breaking algorithm is called only when all the ancillary qubits are assigned to case 4.}.
In such a case, we list ancillary qubits in case 4, i.e., those do not interact with any data qubit after the last Pauli-$Z$ measurement (${\rm CP}(a_i)=\emptyset$). We call such qubits {\it usable}. Then, the tie-breaking algorithm forces moving some usable ancillary qubits so that they become close to a certain data qubit included in the unmeasured stabilizer operators. See Appendix~\ref{app:alg_tie_break} for a detailed explanation and pseudo-code.

\subsection{Working example}
If we obey the above classification rules, the syndrome extraction circuit shown in Fig.\,\ref{fig:managing} is generated. Here, we show a step-by-step explanation on the right of Fig.~\ref{fig:example_cases}.

\begin{itemize}
\item $t=0$: $a_1$ is classified to case 3, and $d_1$ and $d_2$ are the candidates. Thus, a CNOT gate is executed on $a_1$ and $d_1$.
\item $t=1$: $a_1$ is classified to case 1, and only $d_2$ is the candidate. Thus, a CNOT gate is executed on $a_1$ and $d_2$, and $a_1$ is measured since ${\rm CP}(a_1) \in L$.
\item $t=2$: $a_1$ is classified to case 3, and only $d_2$ is the candidate. Thus, a CNOT gate is executed on $a_1$ and $d_2$.
\item $t=3$: $a_1$ is classified to case 2, and $d_2$ is chosen as the neighboring target to shorten the distance to $d_3$. Thus, a SWAP gate is performed on $a_1$ and $d_2$. 
\item $t=4$, $a_1$ is classified to case 1, and only $d_3$ is the candidate. Thus, a CNOT gate is executed on $a_1$ and $d_3$, and $a_1$ is measured since ${\rm CP}(a_1) \in L$. Now $L$ becomes empty, and the algorithm is completed.
\end{itemize}

\subsection{Unnecessary Gate Removal (Algorithm\,\ref{alg:remove_gates})}
While we can obtain syndrome-extraction circuits by running the explained procedure, unnecessary two-qubit gates might be included in it.
For example, if the matrix $M$ has any 1-element in the end, unnecessary CNOT gates have been performed for a single syndrome measurement. Since such gates will induce undesirable errors, we should remove them from $C$.
The aim of Algorithm\,\ref{alg:remove_gates} is to delete such unnecessary gates.
\begin{algorithm}[H]
\caption{Algorithm for unnecessary gate removal}
\label{alg:remove_gates}
\begin{algorithmic}[]
\Require{$(C, M)$}
    \ForAll{$i \in [1..m]$}
        \State{$r \leftarrow \texttt{popcnt}(\texttt{get\_row}(M,i))$}
        \State{$\texttt{remove\_last\_cnot}(C,i,r)$}
    \EndFor
    \State{$\texttt{remove\_consecutive\_swap}(C)$}
    \State \textbf{return} $C$
\end{algorithmic}
\end{algorithm}
In this algorithm, we get the final parity-check status of ancillary qubit $a_i$ from $M$ ($\texttt{get\_row}$ function), and count the number of meaningless CNOT gates on $a_i$, i.e., the remaining number of 1-elements ($\texttt{popcnt}$ function). Then, we check $C$ in reverse order and sequentially delete the CNOT gates that generate the remaining 1-elements in the final $M$ (\texttt{remove\_last\_cnot} function).
After removing some CNOT gates, several consecutive SWAP gates on the same qubit pair may remain. They are deleted from circuit $C$ with \texttt{remove\_consecutive\_swap} function.

\section{Numerical Evaluation}\label{sec:Numerical}
In this section, we numerically evaluate the performance of our method with rotated surface codes on a two-dimensional grid, and demonstrate the trade-off relations between the ratio of ancillary qubits and achievable logical error rates. When we evaluate the performance of distance-$d$ surface code, we repeat the forward and reverse sequences of noisy syndrome measurements for $(d-2)$ rounds, where we assume the initialization and final logical measurement are noiseless.

\begin{figure}[t]
  \centering
  \includegraphics[width=0.5\textwidth]{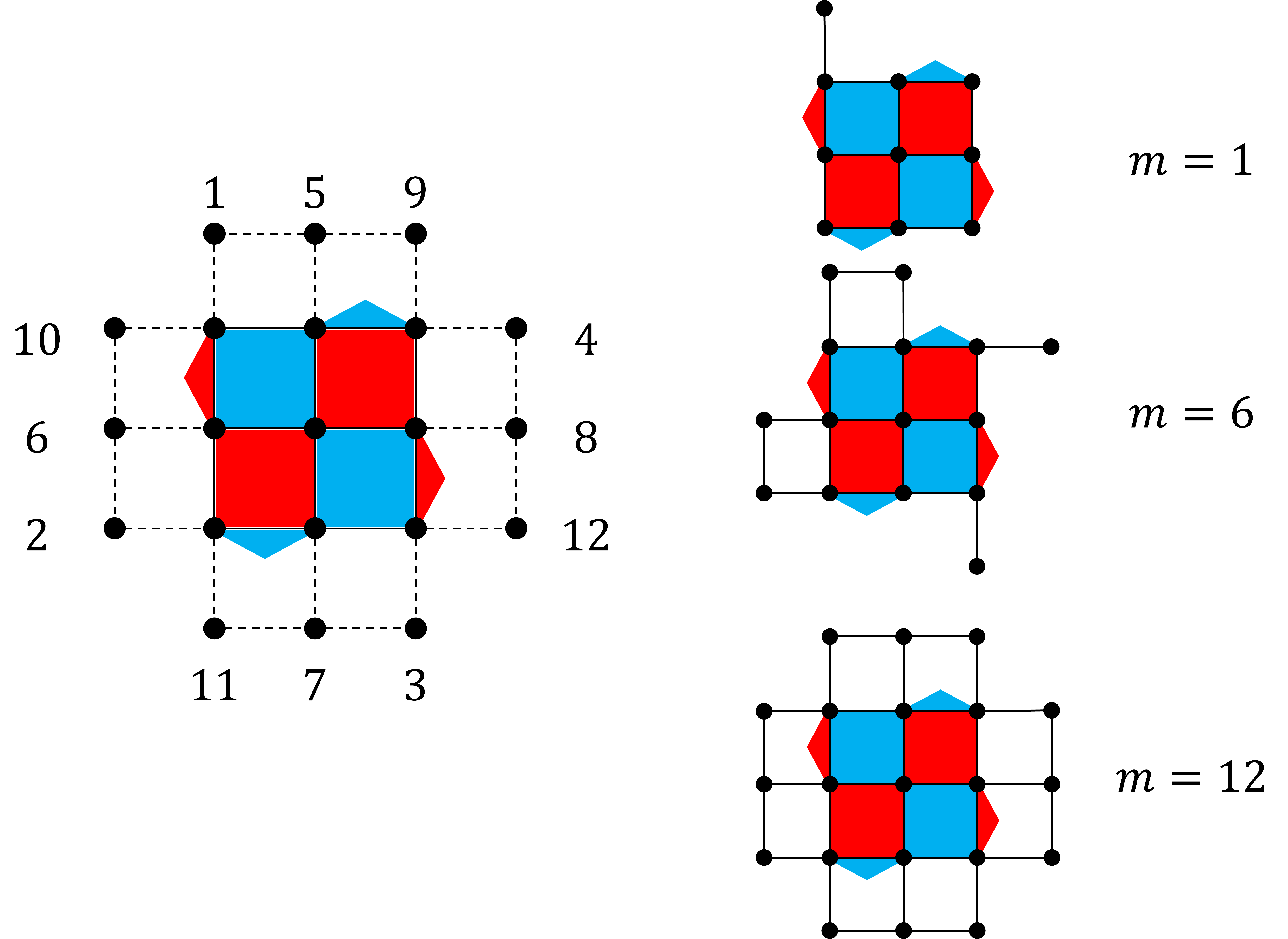}
  \caption{Qubit-connectivity graphs and initial positions with ancillary qubits for distance-$3$ surface codes (left) and several examples with $m=1$ (top right), $m=6$ (middle right), $m=12$ (bottom right).}
  \label{fig:layout}
\end{figure}

The initial layout and the connectivity of the graph $G$ are designed by regularly arranging ancillary qubits surrounding the data qubits. Examples of distance-$3$ surface codes are shown in Fig.\,~\ref{fig:layout}. Here, blue~(red) tiles correspond to the $X\  (Z)$-type stabilizers in the initial data-qubit layout. Note that, unlike standard rotated surface code layout and connectivity, ancillary qubits are not located at the center of blue and red faces. This modification enables us to observe trade-offs between the number of data and ancillary qubits. We chose surface codes for benchmarking to leverage fast decoders, well-established error-estimation platforms, and tunability of code distances, rather than to improve surface codes themselves. Finding good initial positions for given quantum error-correcting codes is left for future work.

In the numerical experiments, we used the following circuit-level noise model. 
\begin{itemize}
    \item Two-qubit depolarizing noise with rate $p_{\rm CNOT}$ is applied after each CNOT gate.
    \item Two-qubit depolarizing noise with rate $p_{\rm SWAP}$ is applied after each SWAP gate.
    \item One-qubit depolarizing noise with rate $p_{\rm idling}$ is applied after each idling gate.
\end{itemize}
We converted the solutions of our scheduling to noisy syndrome-extraction circuits of Stim~\cite{gidney2021stim}, and estimated errors with  PyMatching~\cite{higgott2025sparse},

\subsection{Properties of syndrome measurement circuit}
\begin{figure*}[htbp]
    \centering
    \begin{tabular}{@{}c@{}c@{}c@{}}
        \includegraphics[width=0.33\textwidth]{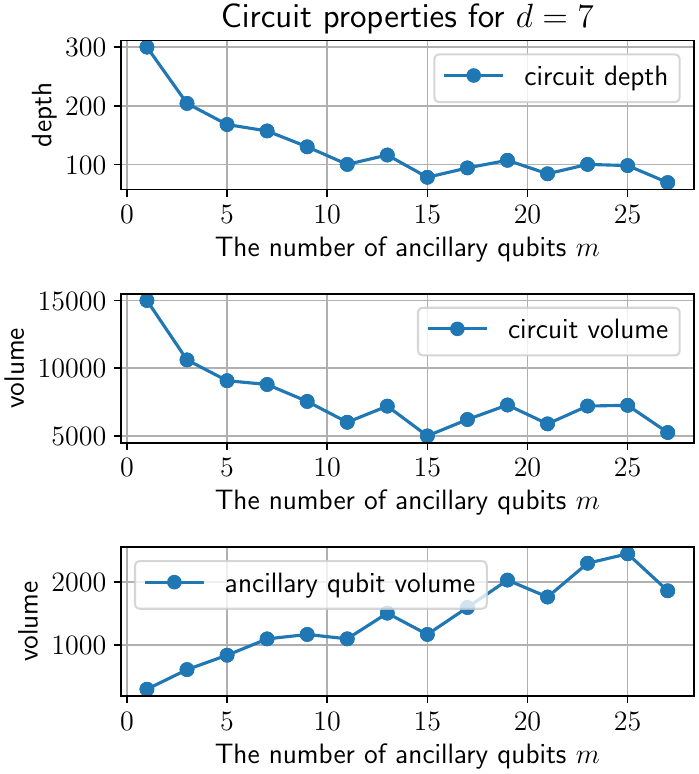} &
        \includegraphics[width=0.33\textwidth]{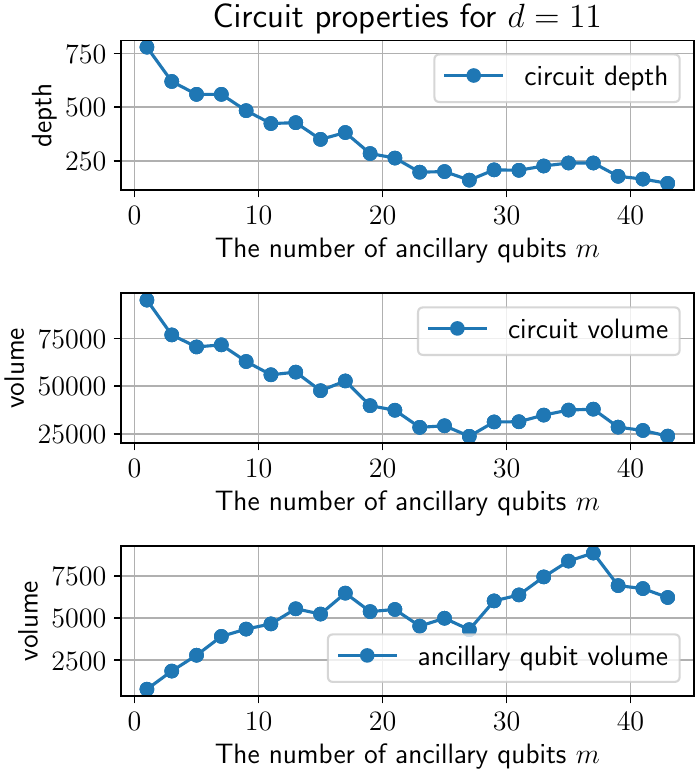} &
        \includegraphics[width=0.33\textwidth]{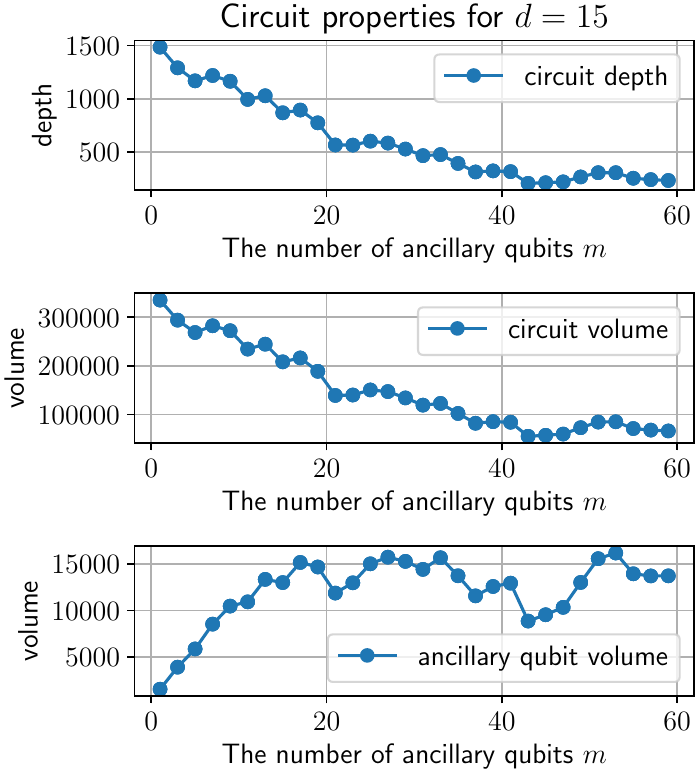} \\
    \end{tabular}
    \caption{The circuit depth, circuit volume, and ancillary circuit volume are plotted in the upper, middle, and bottom rows, respectively. The column corresponds to code distance with $d=7$ (left), $d=11$ (center), and $d=15$ (right).}
    \label{fig:circuit_property}
\end{figure*}

Before estimating logical error rates, we evaluated the properties of obtained solutions with the following three measures: circuit depth, circuit volume, and ancillary qubit volume.
Circuit depth is the length of a critical path of two-qubit gates for a single round of syndrome measurement $S_Z$, which is equal to the step count required for syndrome extraction and is proportional to the number of error events per physical qubit.
Circuit volume is defined as the product of circuit depth and the total number of physical qubits. This value is proportional to the number of error events in each syndrome-extraction cycle.
Ancillary qubit volume is defined as the product of circuit depth and the number of ancillary qubits. If the syndrome measurements are ideally parallelized, the circuit depth is expected to be inversely proportional to the number of ancillary qubits. Thus, the trend of ancillary qubit volume represents how syndrome measurements are parallelized with multiple ancillary qubits.

Figure~\ref{fig:circuit_property} shows the circuit depth, circuit volume, and ancillary qubit volume for the Pauli-$Z$ syndrome measurement in code distance $d=7,11,15$ by varying the number of ancillary qubits.
We confirm that as the number of ancillary qubits increases, the number of circuit depths and circuit volumes decreases. These results suggest that our algorithm can effectively utilize ancillary qubits to generate shallower circuits. Since this will reduce the number of idling error events during the syndrome extraction, it will have positive effects on the logical error rates.
Regarding the ancillary qubit volume, it increases as the number of ancillary qubits increases, and it remains almost constant after a certain point. This means the circuit depth grows faster than the inverse of ancillary qubit counts, and thus indicates that quantum circuits utilize ancillary qubits for achieving better parallelism when the number of ancillary qubits is small. This behavior is expected as follows. In our initial positions of data and ancillary qubits, the ancillary qubits are close to each other. Thus, when the number of ancillary qubits is large, conflicts between ancillary qubits for approaching target data qubits are not negligible. In contrast, such a conflict would be rare in the regime of small ancillary qubit count. This issue negatively affects the logical error rates.

\subsection{Performance under specific gate noise}
We evaluated the dependency of logical error rates on the number of ancillary qubits. 
We checked the logical error rates of surface codes with code distance $d=7,11,15$, and varied the number of ancillary qubits from $1$ to $4d-1$ for distance-$d$ surface codes.
To evaluate the contribution of each gate error, we activate one among the CNOT, SWAP, and idling gate errors.

\begin{figure*}[t]
    \centering
    \makebox[\textwidth][c]{\includegraphics[width=1.0\textwidth]{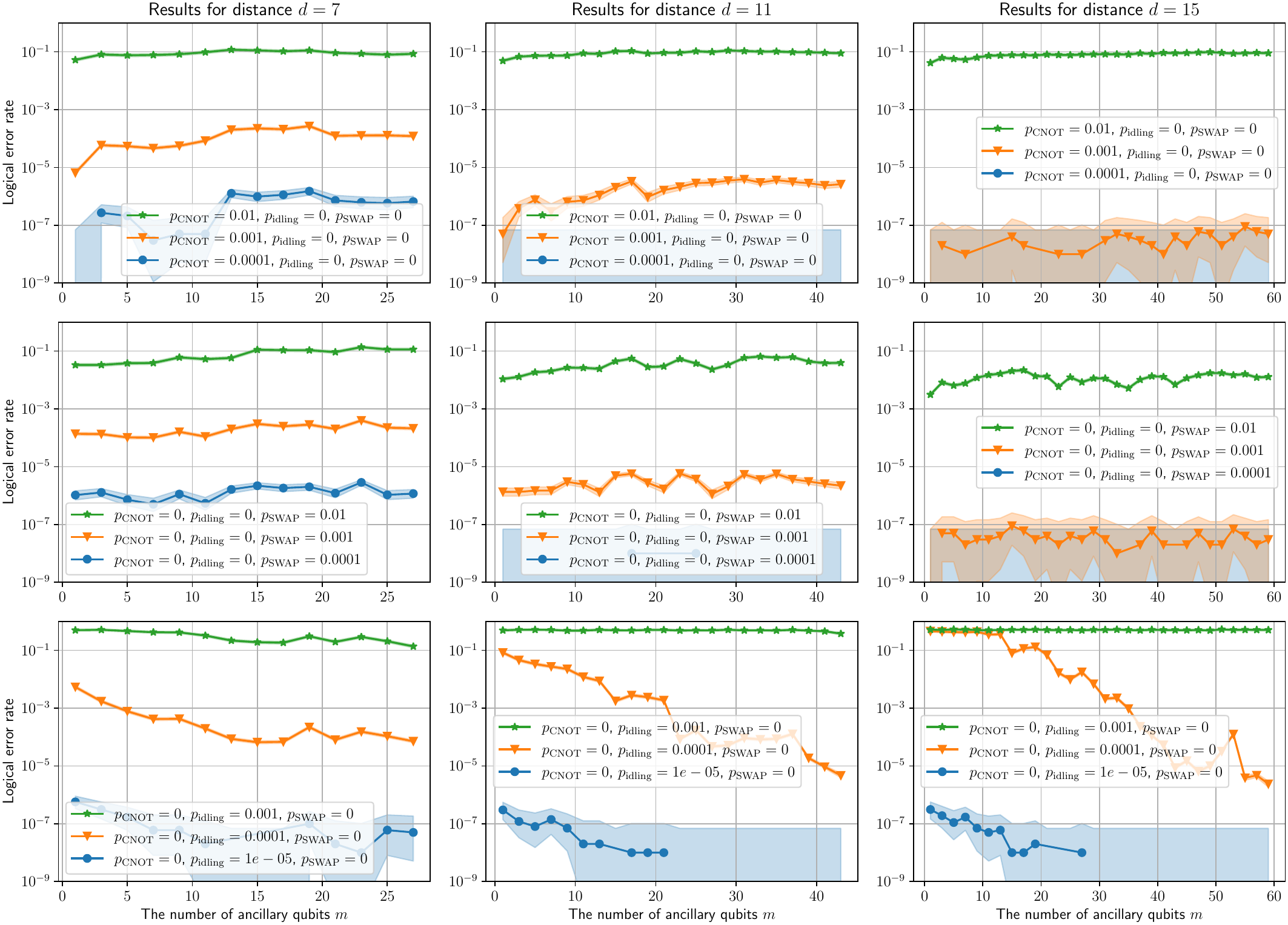}}
    \caption{The logical error rate for distance-$7$ (left), $11$ (center), and $15$ (right) surface codes under CNOT gate errors (upper), SWAP gate errors (middle), and idling errors (bottom).}
    \label{fig:logical_error_surround}
\end{figure*}

The numerical results are shown in Fig.~\ref{fig:logical_error_surround}.
When only CNOT or SWAP gates suffer from errors, we observed that the logical error rates are almost independent of the number of ancillary qubits.
This is a natural consequence because while the lack of ancillary qubits will increase the circuit depth, the number of CNOT gates is constant, and that of SWAP gates will not change significantly. Therefore, the increase in circuit depth does not impose a penalty on the logical error rates.
In contrast, the bottom row indicates that logical error rates increase as the number of ancillary qubits decreases when idling gates only suffer from noise. 

From these experiments, we can conclude that the reduction of ancillary qubits will increase the effect of idling errors while keeping the amount of two-qubit gate errors almost constant. 
Therefore, when the contribution of idling errors to logical error rates is negligible compared to that of two-qubit gate errors, we can reduce the number of ancillary qubits with modest effects on the logical error rates. Such a situation might happen when qubit error rates are dominated by control errors rather than coherence time in neutral atoms and trapped ions~\cite{bluvstein2024logical,reichardt2024demonstration}.

\begin{figure}[t]
    \includegraphics[width=0.5\textwidth]{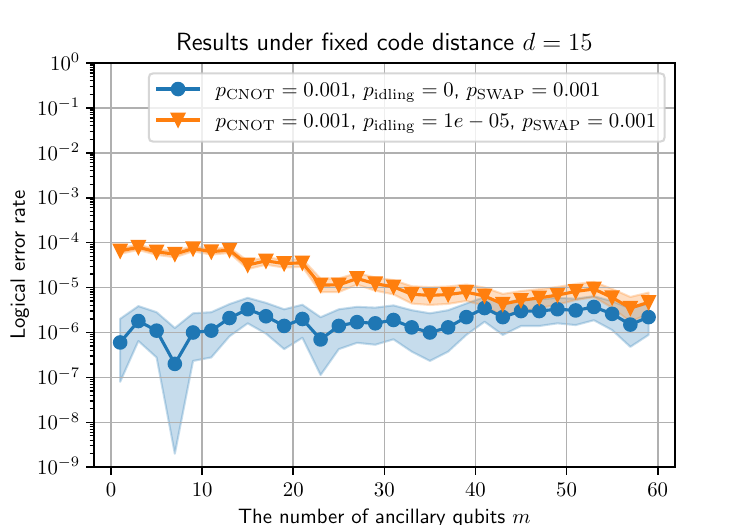}
    \caption{The logical error rate for a fixed code distance.}
    \label{fig:mixed_logical_error_surround}
\end{figure}

To validate this hypothesis, we performed the following experiment. In this experiment, we evaluated the performance with two noise models in a distance-$15$ surface code. 
One assumes errors only occur after each CNOT and SWAP gate, and the other considers all the types of gate errors, including idling gate errors.
The results are shown in Fig.~\ref{fig:mixed_logical_error_surround}.
We confirmed that logical error rates are gradually increasing in the number of ancillary qubits in the first noise model, but are significantly reduced in the latter noise model. 
We expect that the gradual increase of logical error rates in the first noise model happens because scheduling with many ancillary qubits might slightly increase the number of SWAP gates when an ancillary qubit might overtake the other during the syndrome extraction.
This result supports our hypothesis that the penalty of reducing the number of ancillary qubits is dominated by the idling error rate relative to the CNOT and SWAP gate errors.

\subsection{Optimal number of ancillary qubits}
Finally, we fixed the total number of available physical qubits and evaluated the logical error rates by varying the code distance $d$. We fixed the number of physical qubits as $n+m=1000$, varied $d$ as odd numbers between $7$ and $31$. This means the number of physical qubits is $n=d^2$, and residual $1000-d^2$ qubits can be used as ancillary qubits. We varied the number of ancillary qubits and found that too many ancillary qubits do not contribute to reducing logical error rates, and set a heuristic upper-bound as $\min (1000-d^2,4d)$. 
Since we are interested in the cases where the idling errors are smaller than two-qubit gate errors, we choose parameters $p_{\rm CNOT} = p_{\rm SWAP} = 10^{-3}, p_{\rm idling}= 10^{-5}$ for this evaluation.

\begin{figure}[t]
        \includegraphics[width=0.5\textwidth]{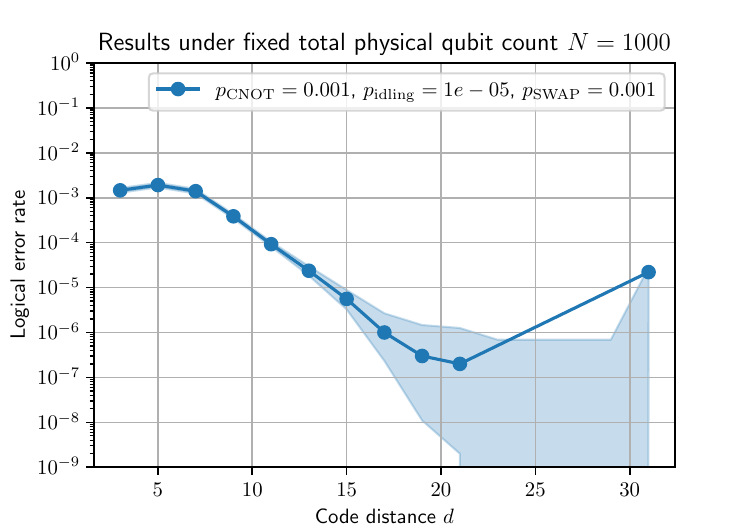}
    \caption{The logical error rate for a fixed total number of available physical qubits.}
    \label{fig:optimal_distance}
\end{figure}

The results are shown in Fig.~\ref{fig:optimal_distance}. 
Note that logical error rates between $d=23$ and $d=29$ are not shown because their logical error rates are too small.
In the region where the code distance is small to medium, there are sufficient ancillary qubits for syndrome measurements, so logical error rates decrease exponentially with the code distance.
On the other hand, when a large number of physical qubits are required for encoding the logical qubit (e.g., $d \sim 31$), the number of ancillary qubits becomes insufficient, which leads to deep syndrome measurement circuits and a significant increase in logical error rates.

This result illustrates that, even in the presence of idling errors, we should not stick to maximized code distance with minimized ancillary qubit count or shallow syndrome measurement circuits using many ancillary qubits. We can obtain the optimal performance by choosing an appropriate number of ancillary qubits and code distance, in particular when the idling error rate is smaller than the two-qubit gate errors.

\section{Conclusion}\label{sec:Conclusion}
In this work, we proposed a framework for generating efficient syndrome measurement circuits using a small number of ancillary qubits for general CSS codes defined on arbitrary connected graphs. We then presented a concrete algorithm for constructing such efficient circuits.
We numerically investigated the impact of various physical noise models on the logical error rates when using a limited number of ancillary qubits. Our results show that increasing the number of ancillary qubits can mitigate the effects of idling errors. In contrast, the impact of two-qubit gate errors on logical error rates is almost independent of the number of ancillary qubits.
Furthermore, we revealed a trade-off between using more ancillary qubits to achieve shallower syndrome-extraction circuits and increasing the number of data qubits to realize larger code distances. Our optimization results indicate that a balanced increase in both resources leads to minimal logical error rates, suggesting a novel design strategy for syndrome measurement circuits, which would be particularly advantageous for qubits with long coherence times.

Since our paper focuses on showing the basic framework and demonstrating trade-off relations, there are various future directions for this research as follows.

\textit{Optimize initial positions:} In this paper, we use a specific graph and initial positions, and there is room for improving logical error rates by optimizing them. While we expect that a shallower circuit can be obtained by putting ancillary qubits among data qubits, it remains unclear how to find the optimal arrangement.

\textit{Extend frameworks:} Our framework focuses on simplified cases to guarantee the halting property of the proposed algorithm. On the other hand, our framework can be straightforwardly extended to allow a wider range of operations. For example, CNOT gates between ancillary qubits allow more efficient collection of error parities. While our framework focuses on CSS codes and stabilizer measurements for Pauli-$X$ and Pauli-$Z$ operators that are repeated independently, we might extend our framework to non-CSS codes and allow simultaneous measurements of stabilizer operators by extending the matrix $M$ to the stabilizer tableau~\cite{gidney2021stim}. 
Also, while our framework assumes that physical qubits are used for data or ancillary qubits, it might be possible to assign residual qubits for detecting flag information~\cite{Yoder2017surfacecodetwist,PhysRevLett.121.050502,Chamberland2018flagfaulttolerant,PRXQuantum.1.010302,PhysRevX.10.011022} to improve the logical error rates.

\textit{Adapt hardware properties:} While we count CNOT gates and SWAP gates independently and equally, some qubit devices have different latencies for each gate, and some devices allow us to execute the combination of them with a shorter pulse sequence. While we ignored the latencies for single-qubit unitary and measurement, extending our framework to treat them would also be straightforward.  Some qubits, such as neutral atoms, can be more flexibly shuttled by optical tweezers. Leveraging these properties would further improve the achievable logical error rates in this framework.

\textit{Optimize algorithms:} We proposed a scheduling algorithm that greedily determines the gate sequence and is completed within a finite time. While finding an optimal scheduling seems computationally hard, we can employ several heuristic methods to reduce the depth of circuits and improve the logical error rates. It is open how these heuristic optimizations can be introduced while keeping halting properties.

\onecolumn\newpage
\appendix

\section{Detail of tie-breaking algorithms}\label{app:alg_tie_break}
This section explains the details of the tie-breaking algorithm as Algorithm~\ref{alg:tie_brekaing}.
This algorithm consists of two main parts. 
First, it searches for pairs of an ancillary qubit and data qubit such that the data qubit is close to the ancillary qubit and its label is in a certain unmeasured labels $l \in L$. In the latter part, SWAP gates are selected to shorten the distance between the ancillary qubit and the data qubit.

\paragraph{Variable definitions}
We first define the set of usable ancillary qubits as the list of ancillary qubits that do not interact with any data qubit after the last measurement.
\begin{align}
    X = \{ a_i\ (i=1,\ldots,m)| {\rm CP}(a_i)=\emptyset \}
\end{align}
We also define the following values.
\begin{align}
    \tilde{D}(l) &= \min_{(d,a) \in l \times X}{\rm dist}(d,a) \\
    (\tilde{d}(l), \tilde{a}(l))&= \mathop{\arg\min}_{(d,a) \in l \times X} {\rm dist}(d,a)
\end{align}
Here ${\rm dist}(\cdot,\cdot)$ is the distance of provided qubits in the graph $G$. Thus, the $\tilde{D}(l)$ represents the shortest distance between the nearest pair with a data qubit from $l$ and an ancillary qubit from the usable ancillary qubits $X$, and a pair $(\tilde{d}(l), \tilde{a}(l))$ is a pair of qubits that achieve the shortest distance.

We also define neighboring path vertices $V^*$ as follows. Let $U$ be a list of the unused qubits, i.e., qubits that are assigned no action. We also denote $P^{-1}$ as the inverse map of qubit positions $P$, i.e., a map from a node in the qubit-connectivity graph to the label of the qubit at the node. Then, for a pair of qubits $q_1, q_2 \in (d_1, ..., d_n, a_1, ... ,a_m)$ the set of vertices $V^*(q_1, q_2, U)$ is determined by 
\begin{align}
    V^*(q_1,q_2,U) = \{ v \in V | (v, P(q_1))\in E, \ {\rm dist}(P^{-1}(v),q_2)<{\rm dist}(q_1,q_2), \ P^{-1}(v) \in U\}.
\end{align}
The first condition requests that the qubit at the node $v$ must be neighboring to $q_1$. The second condition demands that the distance between the node $v$ and that of $q_2$ must be shorter than the distance between $q_1$ and $q_2$, which means the distance between $q_1$ and $q_2$ becomes smaller if $q_1$ is swapped to the qubit at $v$. The final condition demands that no action has been assigned to the qubit in the node $v$ to ensure that we can assign a SWAP gate between $q_1$ and the qubit at $v$.

\paragraph{Algorithm description}
Algorithm\,\ref{alg:tie_brekaing} shows the detailed description of the tie-breaking algorithm.
\begin{algorithm}[H]
    \caption{Tie-breaking algorithm}
    \label{alg:tie_brekaing}
    \begin{algorithmic}[]
    \Require{$(M,P,L,G)$}
    \State{$C_t \leftarrow ()$}
    \ForAll{$q \in (d_1 ... d_n, a_1 ... a_m)$}
        \State{$U[q] \leftarrow \texttt{true}$}
        \Comment{\texttt{true} if qubit unused}
    \EndFor
    \State{$X \gets \texttt{usable\_ancillary\_qubit}(M)$}

    \State{$\texttt{SWAP\_candiates} \gets [\ ]$}
    \ForAll{$l$ in $L$}
        \State{$\tilde{D},\tilde{d},\tilde{a} \leftarrow \texttt{get\_shortest\_distance\_and\_pair}(X,l,P,G)$}
        \State{$\texttt{SWAP\_candiates} \leftarrow \texttt{SWAP\_candiates} + [(\tilde{D}, \tilde{d}, \tilde{a})]$}
    \EndFor
    \State{$\texttt{sort\_with\_distance}(\texttt{SWAP\_candidates})$}
    \ForAll{$(\_, \tilde{d}, \tilde{a}) \in \texttt{SWAP\_candiates}$}
        \State{$V^* \leftarrow \texttt{get\_nehgiboring\_path\_vertices}(\tilde{a}, \tilde{d}, U)$}
        \If{$|V^*| \ge 1 \ {\rm and} \ U[\tilde{a}]=\texttt{true}$}
            \State{$\texttt{swap}(\tilde{a}, V^*[0] ,C_t,P,U)$}
        \EndIf
        \State{$V^* \leftarrow \texttt{get\_nehgiboring\_path\_vertices}(\tilde{d}, \tilde{a}, U)$}
        \If{$|V^*| \ge 1 \ {\rm and} \ U[\tilde{d}]=\texttt{true}$}
            \State{$\texttt{swap}(\tilde{d}, V^*[0] ,C_t,P,U)$}
        \EndIf
    \EndFor
    \State \textbf{return} $C_t, P$
    \end{algorithmic}
\end{algorithm}
We first mark all the physical qubits \texttt{unused} with variable $U$, and enumerate usable ancillary qubits with variable $X$. Then, we list the candidates for SWAP targets to \texttt{SWAP\_candidates}, which is the list of tuples of $(\tilde{D}(l),\tilde{d}(l),\tilde{a}(l))$ for all the $l \in L$. Then, we sort the list of SWAP candidates in ascending order by the distance $\tilde{D}(l)$ and iterate through them for assigning SWAP gates. In each candidate tuple, we try to perform a SWAP gate on the pair of the data qubit and its neighbor, or the pair of the ancillary qubit and its neighbor. To this end, we enumerate the locations of qubits that we can perform SWAP gates to $\tilde{a}$ to make the distance between $\tilde{a}$ and $\tilde{d}$ closer with the function \texttt{get\_nehgiboring\_path\_vertices}, and perform SWAP gates if there is a qubit that satisfies the conditions. If there are multiple neighbors, we choose the one with the smallest index ($V^*[0]$ in the pseudo-code). Then, we do the same procedure by replacing the roles of $\tilde{a}$ and $\tilde{d}$.

\section{Halting property of Algorithm~\ref{alg:main_alg}}\label{app:halting}

In this section, we show that our algorithm always outputs syndrome extraction circuits within a finite time, i.e., it does not enter an infinite loop. A strategy to show this property is as follows.
At each timestep $t$, all the ancillary qubits are classified into case 1, 2, 3, or 4 in Algorithm\,\ref{alg:classify}. We describe the classification result of all the ancillary qubits with four Boolean variables $\mathcal{S}(t) = (\mathcal{S}_1(t),\mathcal{S}_2(t),\mathcal{S}_3(t),\mathcal{S}_4(t)) \in \{\texttt{true}, \texttt{false}\}^4$, where $\mathcal{S}_i(t)$ becomes \texttt{true} if there is at least one ancillary qubit that is assigned to case $i$, and becomes \texttt{false} otherwise. For example, $\mathcal{S}(t) = (\texttt{false}, \texttt{false}, \texttt{false}, \texttt{true})$ represents all the ancillary qubits are assigned to case 4, and $\mathcal{S}(t) = (\texttt{true}, \texttt{true}, \texttt{true}, \texttt{true})$ represents there is at least one ancillary qubit for every case.
Since ancillary qubits are classified into one of cases 1, 2, 3, and 4, at least one among $\mathcal{S}(t)$ is \texttt{true}, and there are $2^{4}-1=15$ possible patterns. 

We will show that these 15 patterns can be classified into four types of situations (Situation A, B, C, D). Then, we show that several repeats of Situation A will decrease the number of unmeasured stabilizer labels $L$, and Situation A must happen after several repetitions of Situation B, C, and D.

\paragraph{Definitions of Situations}
The 15 possible patterns characterized by $\mathcal{S}(t)$ can be classified into the following four types.
\begin{itemize}
\item (Situation A) $\mathcal{S}(t) = (\texttt{true},\texttt{*},\texttt{*},\texttt{*})$ ($8$ patterns), where \texttt{*} represents either of \texttt{true} or \texttt{false}. In this situation type, at least one ancillary qubit is classified into case 1.
\item (Situation B) $\mathcal{S}(t) = (\texttt{false},\texttt{false},\texttt{false},\texttt{true})$ ($1$ pattern). In this situation type, no action is assigned to all the ancillary qubits by Algorithm\,\ref{alg:classify}, and the tie-breaking algorithm is called. 
\item (Situation C) $\mathcal{S}(t) = (\texttt{false},\texttt{true},\texttt{false},\texttt{false})$ or $\mathcal{S}(t) = (\texttt{false},\texttt{true},\texttt{false},\texttt{true})$ ($2$ patterns). In this situation, at least one ancillary qubit is classified into case 2, and it moves towards the target data qubit.
\item (Situation D) $\mathcal{S}(t) = (\texttt{false},\texttt{*},\texttt{true},\texttt{*})$ ($4$ patterns). In this situation type, there is at least one ancillary qubit classified into case 3.
\end{itemize}

When situation A happens, at least one ancillary qubit is assigned to case 1, and the number of data qubits interacting with ancillary qubits (i.e., ${\rm CP}(a)$) will increase. Since the number of qubits is finite, several repetitions of Situation A always result in the completion of stabilizer measurements (i.e., ${\rm CP}(a) = l$ for $l \in L$) and decrease the size of $L$. Therefore, the algorithm halts within a finite time if Situation A occurs regularly. In the discussion below, we will show that situations B, C, and D will eventually transition to situation A. In other words, we will show that an infinite loop among Situation B, C, and D never happens.

{\it Properties of Situation B:}
In the case of situation B, the tie-breaking algorithm (Algorithm~\ref{alg:tie_brekaing}) is executed. The tie-breaking algorithm guarantees that at least one ancillary qubit approaches a fixed data qubit. After the finite step $\tau$, the situation transitions to Situation $D$ as $\mathcal{S}(t+\tau) = (\texttt{false},\texttt{false},\texttt{true},\texttt{true})$, or other situations (Situation A or C) before that. Note that since ancillary qubits with case 1, 2, or 3 do not become case 4 without Situation A, once situation transitions to those other than Situation B, Situation B never happens before Situation A happens.

{\it Properties of Situation C:}
In this situation, there is at least one ancillary qubit that is assigned to case 2, i.e., there is an ancillary qubit $a$ such that ${\rm CP}(a) \neq \emptyset$ but does not have neighboring target data qubits. In this case, ancillary qubits try to approach a certain target data qubit to increase the size of ${\rm CP}(a)$. However, there can be a situation where another ancillary qubit swaps target data qubits, and the distance between the ancillary qubit and the data qubit does not decrease. In this case, an infinite loop can happen inside situation C.
To avoid such an infinite loop, we implement special treatment for the ancillary qubits in case 2 with the smallest index. Let the case 2 ancillary qubits with the smallest index be $a$, and the target data qubit assigned to $a$ be $d$. As shown in Algorithm\,\ref{alg:classify}, the data qubit $d$ is marked as \texttt{used} using the $\texttt{flag}$ variable, and the position of $d$ does not change during Situation C is repeated. Thus, the ancillary qubit $a$ neighbors to $d$ within a finite step, and the situation transitions to Situation A. Note that while the case 2 ancillary qubit with this special treatment might be changed to one with a smaller index, this change happens at most a finite time. 

{\it Properties of Situation D:}
In the situation D, there are at least one ancillary qubit that is assigned to case 3, i.e., an ancillary qubit $a$ that has ${\rm CP}(a)=\emptyset$ but there is a neighboring data qubit $d$ that satisfies $\{d\} \in l$ for a certain $l \in L$.
Let the number of ancillary qubits that does not interact with any data qubit after the last measurement as $m_{\rm usable}$, i.e., $m_{\rm usable} = |\{a \in \{a_1, ..., a_m\} | {\rm CP}(a)=\emptyset\}|$. The existence of case 3 ancillary qubits indicates that $m_{\rm usable} > 0$, and Situation D always decreases $m_{\rm usable}$ since CNOT gates are performed on case 3 ancillary qubits and data qubits. Also, $m_{\rm usable}$ never increases unless situation A happens. Therefore, Situation D can happen for a finite times before Situation A happens. 

In conclusion, we can see that our algorithm has the following properties: 
\begin{enumerate}
\item Situation B transitions to Situation A, C, or D within a finite step. 
\item Situation C and D never transition to Situation B before Situation A. 
\item If Situation C is repeated several times, the situation transitions to Situation A.
\item The number of Situation D after the last Situation A is finite. 
\end{enumerate}
These properties indicate that situation A regularly happens within a finite step, and thus the size of $L$ will decrease continuously, which leads to halting properties of our algorithm.

\bibliographystyle{unsrtnat}
\bibliography{reference}

\begin{thebibliography}{20}
\providecommand{\natexlab}[1]{#1}
\providecommand{\url}[1]{\texttt{#1}}
\expandafter\ifx\csname urlstyle\endcsname\relax
  \providecommand{\doi}[1]{doi: #1}\else
  \providecommand{\doi}{doi: \begingroup \urlstyle{rm}\Url}\fi

\bibitem[Shor(1996)]{shor1996fault}
Peter~W Shor.
\newblock Fault-tolerant quantum computation.
\newblock In \emph{Proceedings of 37th conference on foundations of computer science}, pages 56--65. IEEE, 1996.

\bibitem[Knill and Laflamme(1996)]{knill1996concatenated}
Emanuel Knill and Raymond Laflamme.
\newblock Concatenated quantum codes.
\newblock \emph{arXiv preprint quant-ph/9608012}, 1996.

\bibitem[Knill et~al.(1998)Knill, Laflamme, and Zurek]{knill1998resilient}
Emanuel Knill, Raymond Laflamme, and Wojciech~H Zurek.
\newblock Resilient quantum computation: error models and thresholds.
\newblock \emph{Proceedings of the Royal Society of London. Series A: Mathematical, Physical and Engineering Sciences}, 454\penalty0 (1969):\penalty0 365--384, 1998.

\bibitem[Aharonov and Ben-Or(1997)]{aharonov1997fault}
Dorit Aharonov and Michael Ben-Or.
\newblock Fault-tolerant quantum computation with constant error.
\newblock In \emph{Proceedings of the twenty-ninth annual ACM symposium on Theory of computing}, pages 176--188, 1997.

\bibitem[Acharya et~al.(2022)Acharya, Aleiner, Allen, Andersen, Ansmann, Arute, Arya, Asfaw, Atalaya, Babbush, et~al.]{acharya2022suppressing}
Rajeev Acharya, Igor Aleiner, Richard Allen, Trond~I Andersen, Markus Ansmann, Frank Arute, Kunal Arya, Abraham Asfaw, Juan Atalaya, Ryan Babbush, et~al.
\newblock Suppressing quantum errors by scaling a surface code logical qubit.
\newblock \emph{arXiv preprint arXiv:2207.06431}, 2022.

\bibitem[Acharya et~al.(2024)Acharya, Abanin, Aghababaie-Beni, Aleiner, Andersen, Ansmann, Arute, Arya, Asfaw, Astrakhantsev, et~al.]{acharya2024quantum}
Rajeev Acharya, Dmitry~A. Abanin, Laleh Aghababaie-Beni, Igor Aleiner, Trond~I. Andersen, Markus Ansmann, Frank Arute, Kunal Arya, Abraham Asfaw, Nikita Astrakhantsev, et~al.
\newblock Quantum error correction below the surface code threshold.
\newblock \emph{Nature}, December 2024.
\newblock ISSN 1476-4687.
\newblock \doi{10.1038/s41586-024-08449-y}.
\newblock URL \url{http://dx.doi.org/10.1038/s41586-024-08449-y}.

\bibitem[Bluvstein et~al.(2024)Bluvstein, Evered, Geim, Li, Zhou, Manovitz, Ebadi, Cain, Kalinowski, Hangleiter, et~al.]{bluvstein2024logical}
Dolev Bluvstein, Simon~J Evered, Alexandra~A Geim, Sophie~H Li, Hengyun Zhou, Tom Manovitz, Sepehr Ebadi, Madelyn Cain, Marcin Kalinowski, Dominik Hangleiter, et~al.
\newblock Logical quantum processor based on reconfigurable atom arrays.
\newblock \emph{Nature}, 626\penalty0 (7997):\penalty0 58--65, 2024.

\bibitem[Reichardt et~al.(2024)Reichardt, Aasen, Chao, Chernoguzov, van Dam, Gaebler, Gresh, Lucchetti, Mills, Moses, et~al.]{reichardt2024demonstration}
Ben~W Reichardt, David Aasen, Rui Chao, Alex Chernoguzov, Wim van Dam, John~P Gaebler, Dan Gresh, Dominic Lucchetti, Michael Mills, Steven~A Moses, et~al.
\newblock Demonstration of quantum computation and error correction with a tesseract code.
\newblock \emph{arXiv preprint arXiv:2409.04628}, 2024.

\bibitem[Fowler et~al.(2012)Fowler, Mariantoni, Martinis, and Cleland]{fowler2012surface}
Austin~G Fowler, Matteo Mariantoni, John~M Martinis, and Andrew~N Cleland.
\newblock Surface codes: Towards practical large-scale quantum computation.
\newblock \emph{Physical Review A}, 86\penalty0 (3):\penalty0 032324, 2012.

\bibitem[Fowler and Gidney(2018)]{fowler2018low}
Austin~G Fowler and Craig Gidney.
\newblock Low overhead quantum computation using lattice surgery.
\newblock \emph{arXiv preprint arXiv:1808.06709}, 2018.

\bibitem[De and Pryadko(2013)]{PhysRevLett.110.070503}
Amrit De and Leonid~P. Pryadko.
\newblock Universal set of scalable dynamically corrected gates for quantum error correction with always-on qubit couplings.
\newblock \emph{Phys. Rev. Lett.}, 110:\penalty0 070503, Feb 2013.
\newblock \doi{10.1103/PhysRevLett.110.070503}.
\newblock URL \url{https://link.aps.org/doi/10.1103/PhysRevLett.110.070503}.

\bibitem[Antipov et~al.(2023)Antipov, Kiktenko, and Fedorov]{PhysRevA.107.032403}
A.~V. Antipov, E.~O. Kiktenko, and A.~K. Fedorov.
\newblock Realizing a class of stabilizer quantum error correction codes using a single ancilla and circular connectivity.
\newblock \emph{Phys. Rev. A}, 107:\penalty0 032403, Mar 2023.
\newblock \doi{10.1103/PhysRevA.107.032403}.
\newblock URL \url{https://link.aps.org/doi/10.1103/PhysRevA.107.032403}.

\bibitem[Ye and Delfosse(2025)]{ye2025quantumerrorcorrectionlong}
Min Ye and Nicolas Delfosse.
\newblock Quantum error correction for long chains of trapped ions, 2025.
\newblock URL \url{https://arxiv.org/abs/2503.22071}.

\bibitem[Gidney(2021)]{gidney2021stim}
Craig Gidney.
\newblock Stim: a fast stabilizer circuit simulator.
\newblock \emph{{Quantum}}, 5:\penalty0 497, July 2021.
\newblock ISSN 2521-327X.
\newblock \doi{10.22331/q-2021-07-06-497}.
\newblock URL \url{https://doi.org/10.22331/q-2021-07-06-497}.

\bibitem[Higgott and Gidney(2025)]{higgott2025sparse}
Oscar Higgott and Craig Gidney.
\newblock Sparse blossom: correcting a million errors per core second with minimum-weight matching.
\newblock \emph{Quantum}, 9:\penalty0 1600, 2025.

\bibitem[Yoder and Kim(2017)]{Yoder2017surfacecodetwist}
Theodore~J. Yoder and Isaac~H. Kim.
\newblock The surface code with a twist.
\newblock \emph{{Quantum}}, 1:\penalty0 2, April 2017.
\newblock ISSN 2521-327X.
\newblock \doi{10.22331/q-2017-04-25-2}.
\newblock URL \url{https://doi.org/10.22331/q-2017-04-25-2}.

\bibitem[Chao and Reichardt(2018)]{PhysRevLett.121.050502}
Rui Chao and Ben~W. Reichardt.
\newblock Quantum error correction with only two extra qubits.
\newblock \emph{Phys. Rev. Lett.}, 121:\penalty0 050502, Aug 2018.
\newblock \doi{10.1103/PhysRevLett.121.050502}.
\newblock URL \url{https://link.aps.org/doi/10.1103/PhysRevLett.121.050502}.

\bibitem[Chamberland and Beverland(2018)]{Chamberland2018flagfaulttolerant}
Christopher Chamberland and Michael~E. Beverland.
\newblock Flag fault-tolerant error correction with arbitrary distance codes.
\newblock \emph{{Quantum}}, 2:\penalty0 53, February 2018.
\newblock ISSN 2521-327X.
\newblock \doi{10.22331/q-2018-02-08-53}.
\newblock URL \url{https://doi.org/10.22331/q-2018-02-08-53}.

\bibitem[Chao and Reichardt(2020)]{PRXQuantum.1.010302}
Rui Chao and Ben~W. Reichardt.
\newblock Flag fault-tolerant error correction for any stabilizer code.
\newblock \emph{PRX Quantum}, 1:\penalty0 010302, Sep 2020.
\newblock \doi{10.1103/PRXQuantum.1.010302}.
\newblock URL \url{https://link.aps.org/doi/10.1103/PRXQuantum.1.010302}.

\bibitem[Chamberland et~al.(2020)Chamberland, Zhu, Yoder, Hertzberg, and Cross]{PhysRevX.10.011022}
Christopher Chamberland, Guanyu Zhu, Theodore~J. Yoder, Jared~B. Hertzberg, and Andrew~W. Cross.
\newblock Topological and subsystem codes on low-degree graphs with flag qubits.
\newblock \emph{Phys. Rev. X}, 10:\penalty0 011022, Jan 2020.
\newblock \doi{10.1103/PhysRevX.10.011022}.
\newblock URL \url{https://link.aps.org/doi/10.1103/PhysRevX.10.011022}.

\end{thebibliography}

\end{document}